\documentclass[proof]{WileyASNA-v1}

\articletype{ORIGINAL ARTICLE}%

\received{xx xxxx xx}
\revised{xx xxxx xx}
\accepted{xx xxxx xx}

\begin{document}

\title{Discovery of eclipses in the cataclysmic variable MASTER OT J232100.42+494614.0}

\author{V. P. Kozhevnikov}

\authormark{KOZHEVNIKOV\textsc{}}

\address{\orgdiv{Astronomical Observatory}, \orgname{Ural Federal University}, \orgaddress{\state{Yekaterinburg}, \country{Russia}}}

\corres{V. P. Kozhevnikov, Ural Federal University, 19 Mira Street, 620002  Yekaterinburg, Russia \email{valery.kozhevnikov@urfu.ru}}


\abstract{I performed photometric observations of the poorly studied cataclysmic variable MASTER OT J232100.42+494614.0 and discovered eclipses. I observed nine eclipses over two groups of closely spaced nights, separated by a 12-month interval. I precisely calculated the orbital period of the system to be $P_{orb}=0.212\,371\,533\pm0.000\,000\,069$ days. For the eclipses, I created an ephemeris that is accurate for a long time span and suitable for analysing changes in the orbital period. The star's brightness outside of the eclipses varied between 15.61$\pm$0.02 and 18.04$\pm$0.05 mag. As the brightness decreased, the eclipses became more profound whereas its width remained unchanged. The average depth of eclipses was 1.36$\pm$0.03~mag, and the average width at half-depth was 0.0512(15) phases. I calculated the possible inclination of the orbit to be between $72.8^\circ$ and $76.3^\circ$, and the range of the average absolute $V$-band magnitude of the disc to be between 6.37$\pm$0.35 and 6.68$\pm$0.35~mag. Although the light curve of MASTER OT J232100.42+494614.0 from the ZTF survey showed insufficiently large outbursts with amplitudes less than 2.5~mag, it should be classified as a dwarf nova, as the average disc brightness and mass transfer rate are below the threshold for thermal instability.}


\jnlcitation{\cname{%
\author{Kozhevnikov V. P.}, 
} (\cyear{2026}), 
\ctitle{Discovery of eclipses in the cataclysmic variable MASTER OT J232100.42+494614.0}, \cjournal{Astron. Nachr.}, \cvol{xxxx;xx:x-x}.}


\maketitle


\section{Introduction}\label{sec1}

Cataclysmic variables (CVs) are binary star systems consisting of a white dwarf and a red dwarf. In these systems, the red dwarf fills its Roche lobe, causing it to transfer material to the white dwarf, creating a bright accretion disc around it. If the white dwarf possesses a strong magnetic field, this disc can be truncated or replaced by a stream, as observed in polar and intermediate polar systems. Due to the asynchronous rotation, some intermediate polars may display coherent short-period oscillations linked to the spin of the white dwarf. All CVs exhibit rapid, non-periodic variations in brightness known as flickering. CVs can be categorized as either dwarf novae or nova-like variables depending on their photometric characteristics. After an eruption, classical novae show similar characteristics to those of nova-like variables. Dwarf novae undergo outbursts that last from 2 to 20 days, with brightness changes ranging from 2 to 5~mag. Nova-like variables either do not experience outbursts or their brightness changes are less than 1~mag. However, some novae-like variables, such as VY Scl stars, can occasionally experience a drop in brightness of more than 1~mag compared to their normal brightness. When the angle between the orbital plane and the observer's line of sight is small, eclipses may occur. Eclipsing CVs provide secure information about the orbital period and the orbital inclination. Comprehensive discussions of CVs can be found in \citet{ladous94}, \citet{warner95} and \citet{hellier01}. More recently, \citet {hardy17} provided up-to-date data on eclipsing CVs.

Many CVs and CV candidates were discovered using MASTER robotic telescopes \citep{lipunov10}. Using the Open Cataclysmic Variable Catalogue (\url{https://depts.washington.edu/catvar/}), I selected from these objects those with declinations greater than $30^\circ$, and which could be observed most of the night at our observatory, located 80~km from Yekaterinburg. There were 337 such objects. Approximately 80\% of them had photometric data from the ZTF survey (\url{https://irsa.ipac.caltech.edu/cgi-bin/Gator/nph-scan?submit=Select&projshort=ZTF}). After reviewing these data, I found that most of these stars are very faint in their normal state (20--22 mag), and their detailed photometry presents a challenge. However, 21 stars were brighter than 18.5~mag for most of the time, and such stars can be centred in the diaphragm of a three-channel photometer, which I use for photometric measurements. My focus has been on the object MASTER OT J232100.42+494614.0 (hereafter referred to as J2321), which was discovered as a new cataclysmic variable (dwarf nova in outburst) by \citet{shumkov14}.

From the light curve of J2321, obtained during the ZTF survey, I learned that its brightness was mostly between 18 and 18.5~mag, with occasional outbursts reaching up to 16~mag. In addition, about a dozen points showed drops to 19.5~mag, suggesting eclipses. However, the International Variable Star Index (VSX) database (\url{https://vsx.aavso.org/index.php?view=search.top}) indicates that eclipses in J2321 have not been detected. I conducted photometric observations to detect eclipses in this star. This paper presents the results obtained.

\section{Observations}\label{sec2}

To observe variable stars, I use a three-channel photometer equipped with photomultipliers. This instrument works by converting photons into electrical pulses, allowing for precise photon counting and light intensity measurements. The design and noise analysis of the photometer are described in \citet{kozhevnikoviz00}. The photometer constantly monitors the light intensity of two stars and the ambient sky background in the field of view of the telescope. This method ensures accuracy even in difficult circumstances, such as fluctuating atmospheric transparency and a variable sky background. Three pulse counters transmit information to a computer. The photometer operates with a 70-cm Cassegrain telescope, which is equipped with step motors controlled by a computer. A CCD-based guiding system is used to compensate for tracking errors and precisely center two stars within the photometer diaphragms during observations. All elements, including the telescope, photometer and guiding system, operate under computer supervision. Human intervention is only needed when thick clouds interfere with the observation process.

I conducted photometric measurements using unfiltered white light, spanning a wavelength range of approximately 3000 to 8000\AA, with a time resolution of 16~s. Because J2321 was expected to be a relatively faint object (18--18.5~mag), I measured the light intensity of two stars using small diaphragms with a diameter of 10~arcsec to reduce the contribution of sky background. Unfortunately, this limited the observation time during the night, as stellar images were less affected by atmospheric turbulence when stars were higher above horizon. Conversely, I used a larger 40-arcsec diaphragm to measure the light intensity of the sky background in the third channel of the photometer. This helps minimize photon noise from Poisson fluctuations of sky background photons.

Fortunately, on the first night of observations, J2321 was quite bright (approximately 16~mag). Although stars with such brightness are not visible to the eye with our 70-cm telescope, the total light intensity from this star and the sky background was significantly higher than the sky background alone, making it possible to locate and centre J2321 within the photometer's diaphragm by moving the telescope according to stellar coordinates and observing photometric measurements. During subsequent observations, the photometer settings were unchanged, allowing me to centre each of the two stars within their respective diaphragms using a comparison star that was clearly visible. 

The comparison star is Gaia DR3 1943112306444903552, which has $G$=11.68 and $BP-RP$=1.59~mag \citep{gaia16, gaia23}. Its colour is similar to that of J2321. According to the coordinates, J2321 corresponds to Gaia DR3 1943116807570569216, which has $G$=18.56 and $BP-RP$=1.22~mag.  The similarity in colours minimizes the effect of differential extinction. Moreover, the differences in photomultiplier sensitivities and the differences in sky background measurements, caused by differences in diaphragm sizes, are taken into account when calculating differential magnitudes.

During the first observation night, I observed an eclipse in J2321. As mentioned, eclipses in J2321 had not been recorded in the VSX database, so I continued to observe this star in order to accurately determine the period of the eclipses. The observations lasted for a total of 31 hours over eight nights and covered 12 months. This information is reflected in the observational log, which is shown in Table~\ref{tab1}.


\begin{table}[t]
\caption{Log of the observations}
\label{tab1}
\begin{tabular}{@{}l c c c}
\hline
\noalign{\smallskip}
Date (UT) & BJD$_{\rm TDB}$ start & Length    & Average out-of-ecl. \\ 
                 & (-2459000)                  &   (h)       &   magnitude \\
\noalign{\smallskip}
\hline
\noalign{\smallskip} 

2022 Sep. 2     &   825.277654     &   1.4   &      15.61$\pm$0.02   \\
2022 Sep. 3     &   826.362679     &   1.2   &      15.72$\pm$0.04   \\
2022 Sep. 7     &   830.289965     &   3.6   &      15.98$\pm$0.03   \\
2022 Sep. 28   &   851.196195     &   7.1   &      17.99$\pm$0.05  \\
2023 Sep. 6     &   1194.286033   &   4.6   &      16.30$\pm$0.02  \\
2023 Sep. 12   &   1200.296632    &   3.6   &     17.91$\pm$0.05  \\
2023 Sep. 13   &   1201.315222    &   3.7   &     18.04$\pm$0.05  \\
2023 Sep. 15   &   1203.226748    &   5.5   &     18.03$\pm$0.05  \\

\noalign{\smallskip}
\hline
\end{tabular} 
\end{table}


\begin{figure}[t]
\centering
\includegraphics[width=84mm]{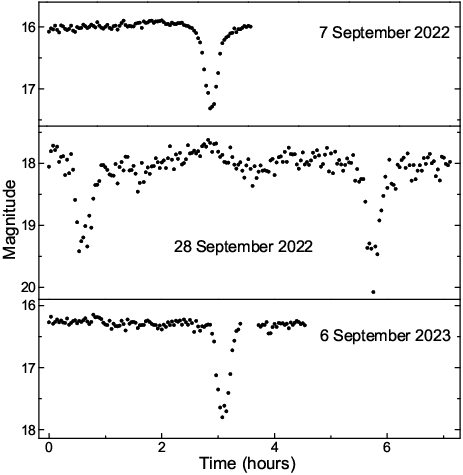}
\caption{Longest light curves of J2321, obtained with a time resolution of 128~s, reveal the presence of eclipses} 
\label{fig1}
\end{figure}

\section{Analysis and results}\label{sec3}

As demonstrated in Table 1, for the purpose of this analysis, I used Barycentric Julian Dates in Barycentric Dynamical Time (BJD$_{\rm TDB}$) as a uniform time system. I converted the recorded UTCs to simple Julian Dates (JD$_{\rm UTC}$) and then transformed them into BJD$_{\rm TDB}$ using an online tool described in \citet{eastman10}, available at  
\url{https://arbiter.nextastro.org/toolkit/bjd-converter}. To validate my calculations, I also used the BARYCEN program from the "aitlib" IDL library at the University of T\"{u}bingen (\url{http://astro.uni-tuebingen.de/software/idl/aitlib/}) to convert UTCs to BJD$_{\rm UTC}$. During the monitoring of J2321, the discrepancy between BJD$_{\rm TDB}$ and BJD$_{\rm UTC}$ remained constant, with BJD$_{\rm TDB}$ being 69.2~s greater than BJD$_{\rm UTC}$. This is consistent with the number of leap seconds plus 32.184~s, which is a fixed difference from International Atomic Time \citep{eastman10}. The accuracy of the time measurements in my observations is within a few tenths of a second.


In Figure~\ref{fig1}, which illustrates the light curves of J2321 featuring eclipses, as well as in other suitable cases, I have converted differential magnitudes to magnitudes using the $G$-band magnitude of the comparison star. The $G$ band used in the $Gaia$ mission is similar to the wavelength range used in my observations with photomultipliers, making it a suitable choice. Furthermore, the colour of the comparison star is similar to that of J2321, with only a difference of 0.37 mag in $BP-RP$, indicating small differences between photometric systems.

The photon counts should follow a Poisson distribution with $N=\sigma^2$. Here, $N$ represents the average number of photons, and $\sigma^2$ is the variance of the data points in the time series. I confirmed this by using a stable light source for my photomultipliers. The result was $N/\sigma^2$=0.98$\pm$0.01, allowing me to estimate the photon noise in the light curves of J2321. To calculate the photon noise for a specific stellar channel, I used a formula provided by \citet{kozhevnikov25}. 

To make eclipses clearer, the light curves shown in Figure~\ref{fig1} were obtained with a time resolution of 128~s. With this time resolution, the photon noise is much less than it was with an initial time resolution of 16~s. As seen, the brightness of J2321 outside of eclipses varies between 16 and 18~mag (see also Table~\ref{tab1}). The photon noise in these light curves outside the eclipses ranges from 0.02 to 0.03~mag when the star is bright, and is 0.06~mag when it is faint. Within the eclipses, the photon noise varies from 0.04 to 0.11~mag. If I used an initial time resolution of 16~s, the photon noise would be higher by a factor of $\sqrt{8}$.


\begin{table}[t]
\caption[ ]{Parameters of the observed eclipses}
\label{tab2}
\begin{tabular}{@{}l c c c}
\hline
\noalign{\smallskip}
Date (UT)  &   Depth      & Width      &  \scriptsize BJD$_{\rm TDB}$ mid-ecl.   \\
                  &  (mag)      &  (min)     & (-2459000)                        \\
\noalign{\smallskip}
\hline
\noalign{\smallskip}

2022 Sep. 2       & 0.890(13)     & 16.75$\pm$0.33    &   825.31325(7)      \\
2022 Sep. 3       & 0.961(21)     & 15.60$\pm$0.45    &   826.37516(10)    \\
2022 Sep. 7       & 1.287(24)     & 15.52$\pm$0.40    &   830.41017(9)      \\
2022 Sep. 28a   & 1.442(103)     & 18.89$\pm$1.80    &   851.22291(42)     \\
2022 Sep. 28b   & 1.704(110)     & 14.09$\pm$1.20    &   851.43554(30)     \\
2023 Sep. 6       & 1.547((43)     & 14.49$\pm$0.50    &  1194.41488(13)    \\
2023 Sep. 12     & 1.277(77)     & 16.58$\pm$1.30    &  1200.36123(33)    \\
2023 Sep. 13     & 1.572(116)     & 15.20$\pm$1.41    &  1201.42379(36)    \\
2023 Sep. 15     & 1.600(128)     & 13.81$\pm$1.41    &  1203.33517(35)    \\

\noalign{\smallskip}
\hline
\end{tabular}
\end{table}


\begin{figure}[t]
\centering
\includegraphics[width=84mm]{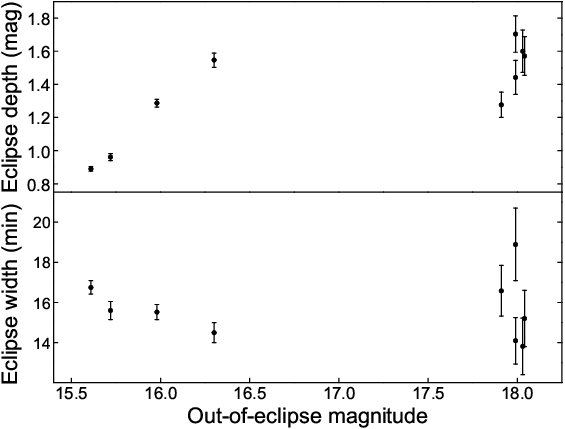}
\caption{Depth (top panel) and width (bottom panel) of the eclipse in relation to the magnitude of J2321 outside the eclipse}
\label{fig2}
\end{figure}

I calculated the average out-of-eclipse magnitude for each night of observation by looking at the light curve graph and averaging the individual measurements taken outside of eclipses. These out-of-eclipse magnitudes are listed in Table~\ref{tab1} and range from 15.61$\pm$0.02 to 18.04$\pm$0.05~mag. These variations resemble changes during dwarf nova outbursts, although they seem to be insufficiently large.

I used a time resolution of 64~s to determine the eclipse parameters. If I had used a time resolution of 16~s, it would have been impossible to calculate the magnitudes for the deepest points of eclipses due to large fluctuations in the number of recorded photons. On the other hand, the time resolution used in the light curves in Figure~\ref{fig1}, 128~s, seems insufficient because it leads to underestimating errors in determining eclipse parameters due to a small number of points in the eclipse profile. Light curves with a time resolution of 64~s are available for analysis at: \url{https://www.researchgate.net/publication/415561228}. I calculated the depth, width at half-depth, and mid-eclipse time of the observed eclipses by fitting a Gaussian curve to the eclipse profiles. The results of these calculations are summarized in Table~\ref{tab2}.

The average eclipse depth is 1.36$\pm$0.10~mag. As seen from Tables~\ref{tab1} and ~\ref{tab2}, the depth of eclipses increases significantly with the magnitude of J2321 outside of eclipses. The range is from 0.890$\pm$0.013 to 1.704$\pm$0.110~mag, In the top panel of Figure~\ref {fig2}, I plot the depth of the eclipse against the magnitude outside the eclipse. For the first data points, there is a linear correlation between these two quantities over a range of 15.61$\pm$0.02 -- 16.30$\pm$0.02~mag. Specifically, the relationship between them is: eclipse depth(mag) = -14.51$\pm$1.11 + 0.986$\pm$0.070 $\times$ out-of-eclipse magnitude. The statistical significance of this relationship is high, as the slope exceeds 14 times its standard error.   However, there are no significant changes in the depth of eclipses when the magnitude of J2321 outside eclipses varies from 16.30$\pm$0.02 to 17.99$\pm$0.02~mag. In fact, the average depth of the last five eclipses, which are grouped around a magnitude of 17.99$\pm$0.02~mag, is 1.52$\pm$0.07~mag. The fourth eclipse, corresponding to a magnitude of 16.30$\pm$0.02~mag, had a depth of 1.55$\pm$0.04~mag. The difference between the two values is relatively small and does not exceed the standard error.


The average eclipse width at half-depth is 15.7$\pm$0.5~min. In contrast to the depth of eclipses, the width of eclipses does not significantly vary with the magnitude of J2321 outside of eclipses. This s shown in the bottom panel of Figure~\ref{fig2}. Although the first four data points in the range of out-of-eclipse magnitudes from 15.61$\pm$0.02 to 16.30$\pm$0.02~mag appear to show a slight negative trend in eclipse widths, this trend is not statistically significant. The slope of the fitted line is only three times the standard error, indicating that the trend is not reliable. Moreover, all data points do not show any statistically significant trend in eclipse widths.

It is thought that fitting an ephemeris to eclipse times yields the most accurate results for determining the eclipse period. To employ this method, I require a preliminary ephemeris. A pair of successive eclipses seen in Figure~\ref{fig1} enables me to calculate the period directly using the mid-eclipse time of each eclipse. The period calculated from this pair of eclipses is 5.103$\pm$0.012~hr (see Table~\ref{tab2}), but this period is not precise enough. The accumulated error of this period is one cycle in 90 days. The validity time of the ephemeris is approximately four times shorter than the accumulated error (e.g., K. Mukai, \url{https://asd.gsfc.nasa.gov/Koji.Mukai/iphome/iphome.html}). It covers only 20 days, which is much shorter than the duration of observations. A more accurate period is needed to derive a more precise preliminary ephemeris.


\begin{figure}[t]
\centering
\includegraphics[width=84mm]{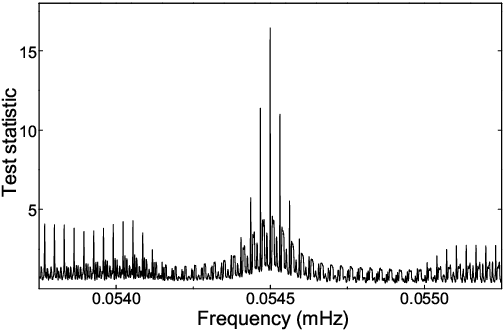}
\caption{AoV spectrum calculated using four light curves from J2321 when it was brighter than 16.30 $\pm$ 0.02 mag outside eclipses}
\label{fig3}
\end{figure}

A more precise period can be determined through the analysis of variance (AoV) method, as suggested by \citet{schwarzenberg89}. For non-sinusoidal signals, such as eclipses, the AoV spectrum is more advantageous than the Fourier power spectrum, which spreads power across numerous high-frequency harmonics \citep{schwarzenberg98}. Unfortunately, there is currently no straightforward method for calculating period errors in AoV spectra. Therefore, I employed the method proposed by \citet{schwarzenberg91} to estimate period errors in Fourier power spectra. However, in my previous research, I discovered that this method consistently overestimated the period error several times when applied to AoV spectra \citep[e.g.,][]{kozhevnikov21}.

\begin{figure}[t]
\centering
\includegraphics[width=84mm]{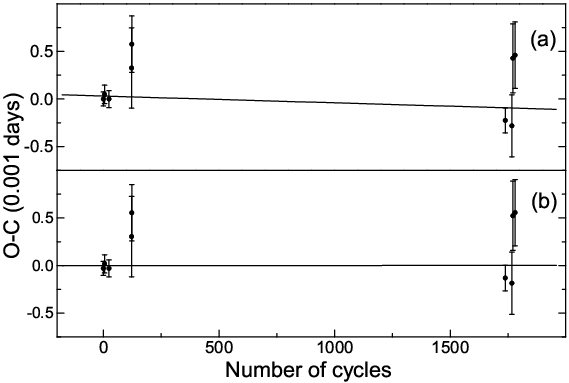}
\caption{(a) The O--C diagram for the preliminary ephemeris shows a noticeable trend. (b) The O--C diagram for the final ephemeris does not show any trend}
\label{fig4}
\end{figure}


\begin{table}[t]
\caption[ ]{Verification of the ephemeris}
\label{tab3}
\begin{tabular}{@{}l c c c}
\noalign{\smallskip}
\hline
\noalign{\smallskip}
Date (UT) & \scriptsize BJD$_{\rm TDB}$ mid-ecl.  & Number   &  O--C              \\
                          & (-2459000)                               & of cycles  & (0.001 days)          \\
\noalign{\smallskip}
\hline
\noalign{\smallskip}

2022 Sep. 2         & 825.31325      & 0        & 0$\pm$0.07   \\
2022 Sep. 3         & 826.37516      & 5        & 0.05$\pm$0.10    \\
2022 Sep. 7         & 830.41017      & 24      & 0.00$\pm$0.09  \\
2022 Sep. 28a     & 851.22291      & 122     & 0.32$\pm$0.42  \\
2022 Sep. 28b     & 851.43554      & 123     & 0.58$\pm$0.30    \\
2023 Sep. 6         & 1194.41488    & 1738   & -0.23$\pm$0.13   \\
2023 Sep. 12       & 1200.36123    & 1766   & -0.28$\pm$0.33   \\
2023 Sep. 13       & 1201.42379    & 1771   & 0.43$\pm$0.36   \\
2023 Sep. 15       & 1203.33517    & 1780   & 0.46$\pm$0.35   \\

\noalign{\smallskip}
\hline
\end{tabular}
\end{table}


As seen from Tables~\ref{tab1} and ~\ref{tab2}, the error in the mid-eclipse time strongly depends on the brightness of J2321, differing by about a factor of 4 for cases where the star is either bright or faint. Unfortunately, my computer program for calculating AoV spectra is not able to use weighted data. Therefore, to find the difference, I calculated the AoV spectrum using all eight light curves and the AoV spectrum only using four light curves when J2321 was brighter than 16.30$\pm$0.02~mag outside eclipses. Both AoV spectra yielded similar periods, but the second AoV spectrum showed a significantly narrower peak, indicating higher accuracy. This second AoV spectrum is presented in Figure~\ref{fig3}. I used a Gaussian fit to the peak to accurately determine its maximum and calculate its error. The resulting period is 0.212\,371\,60(44)~days. This period is accurate enough for a preliminary ephemeris.

To create a preliminary ephemeris, I used the mid-eclipse time of the initial eclipse and the period determined using the AoV spectrum. I then calculated the differences between the observed and calculated (O--C) mid-eclipse times for all subsequent eclipses (see Table~\ref{tab3}). The O--C diagram in Figure~\ref{fig4}a illustrates that error bars vary significantly. This is mostly due to different photon noise, depending on the brightness of J2321. Therefore, for each O--C value, I assigned a weight equal to its error and obtained a linear relationship: $O-C = 0.000\,030(48) - 0.000\,000\,071(69) {\it E}$. As seen in Figure~\ref{fig4}a, this graph exhibits a noticeable slope. After removing this slope and minor vertical displacement, I obtained a final ephemeris without slope or vertical displacement, as shown in Figure~\ref{fig4}b. The final ephemeris is:


{\scriptsize
\begin{equation}
BJD_{\rm TDB}(\rm mid-ecl.) = 245\,9825.313\,284(48)+0.212\,371\,533(69) {\it E}. 
\label{ephemeris1}
\end{equation} }


The final orbital period, determined by fitting the preliminary ephemeris to the O--C values, is 0.212\,371\,533(69) days. This period is close to that obtained from the AoV spectrum, but with an error that is six times less. The overestimation of the error in the AoV spectrum is consistent with my previous research \citep[e.g.,][]{kozhevnikov21}. The accumulated error of the period corresponds to one cycle of oscillation approximately every 1800 years.  Although the validity time of Ephemeris~\ref{ephemeris1} is four times shorter than this time, it remains valid for a very long time up to 450 years.

\begin{figure}[t]
\centering
\includegraphics[width=84mm]{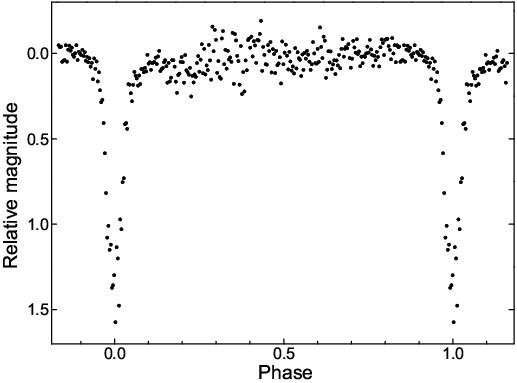}
\caption{Light curves of J2321, folded with the most accurate orbital period}
\label{fig5}
\end{figure}

Figure~\ref{fig5} shows the light curves of J2321, which were folded with the orbital period determined by fitting the preliminary ephemeris. The time between data points is 64~s. As seen in the graph, the scatter of points outside the eclipse is not uniform along the orbital phase. This is due to the short duration of individual light curves of the star in its bright state. These light curves mainly cover only the eclipses and their immediate surroundings. So, the errors of points outside the eclipse range from 0.02 to 0.25~mag, while the average error of points within the eclipse is 0.13~mag. The depth of the eclipse in the folded light curve, as determined by a Gaussian fit, is 1.36$\pm$0.03~mag. This is consistent with the average eclipse depth obtained from individual eclipses, which is the same, except for the error, namely 1.36$\pm$0.10~mag (see Table~\ref{tab2}).


\begin{figure}[t]
\centering
\includegraphics[width=84mm]{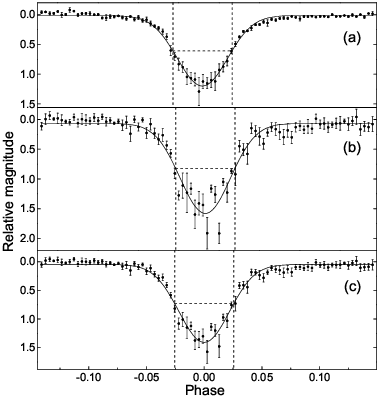}
\caption{Average eclipse profiles depending on the brightness of J2321: The average magnitude outside eclipses was 15.90$\pm$0.28 mag (a) and 17.99$\pm0.03$ mag (b). Frame (c) shows the average eclipse profile obtained by combining all observations. Solid lines represent fitted Gaussian curves, while dashed lines indicate the width of eclipses at half-depth}
\label{fig6}
\end{figure}

As seen in Figure 2, the depth of eclipses varies depending on the brightness of J2321. To investigate how eclipse profiles differ, I classified light curves into two groups: those taken when J2321 was relatively bright (bright state) and those near its minimum brightness (faint state). The first group includes four light curves with out-of-eclipse magnitudes ranging from 15.61$\pm$0.02 to 16.30$\pm$0.02~mag, with an average out-of eclipse magnitude of 15.90$\pm$0.28~mag. The second group consists of four light curves with out-of-eclipse magnitudes between 17.91$\pm$0.05 and 18.04$\pm$0.05~mag, with an average out-of-eclipse magnitude of 17.99$\pm$0.03~mag. There are four eclipses in the first group and five in the second.

I obtained average eclipse profiles by folding the light curves with the orbital period for each group. These profiles, along with the profile obtained from all light curves, are presented in Figure~\ref{fig6}. As seen, the shape of the profiles does not show any significant differences and follows a Gaussian curve. The eclipse profiles only show changes in depth. Using a Gaussian fit, I found that the eclipse depths vary between 1.18$\pm$0.02 and 1.51$\pm$ 0.05~mag for the bright and faint states, respectively. The depth of the profile obtained using all light curves is 1.36$\pm$0.03~mag. These values are consistent with the average depths obtained from individual eclipses, which are 1.17$\pm$0.15, 1.52$\pm$0.07, and 1.36$\pm$0.10~mag for three cases (see Table~\ref{tab2}). In contrast, the eclipse widths at half-depth do not vary significantly. As determined by using a Gaussian fit, the eclipse widths are 0.0513(8) and 0.0513(22) orbital phases for the bright and faint states, respectively. When all light curves are combined, the eclipse width becomes 0.0512(15) orbital phases. Interestingly, the differences between these values are much smaller than their errors. This might mean that the errors calculated using a Gaussian fit are overestimated.  However, these eclipse widths and their errors are consistent with the average eclipse width and the error found using individual eclipses, which is 0.0512(17) orbital phases (see Table~\ref{tab2}). Here the error is determined as the classical RMS error of the average.


To identify coherent short-period oscillations, such as those observed in intermediate polars, I analysed six longest segments of the light curves of J2321, obtained with a time resolution of 16~s that do not contain eclipses. There were two segments corresponding to the bright state of the star and four corresponding to its faint state. These segments had durations ranging from 1.8 to 3.2~hr. To eliminate nightly averages and variations with periods longer than the light curve length, I subtracted a linear fit and then calculated Fourier power spectra for these light curves.

Since the light curves were obtained with varying brightness of the star, the resulting power spectra contain varying levels of photon noise. Consequently, I divided the power spectra into two groups depending on the brightness of J2321. The first group included two power spectra obtained when J2321 was in its bright state with an average magnitude of 16.14$\pm$0.16~mag outside of eclipses. The second group consisted of four power spectra obtained when J2321 was in its faint state with an average magnitude of 17.99$\pm$0.04~mag outside of eclipses. Neither of the power spectra showed any significant peaks up to frequencies of 31~mHz, suggesting that there were no coherent short-period oscillations above the noise level. Noise peaks in the averaged power spectra at frequencies higher than 1~mHz reached amplitudes of about 17~mmag in the bright state of J2321 and 55~mmag in its faint state.


\begin{figure}[t]
\centering
\includegraphics[width=84mm]{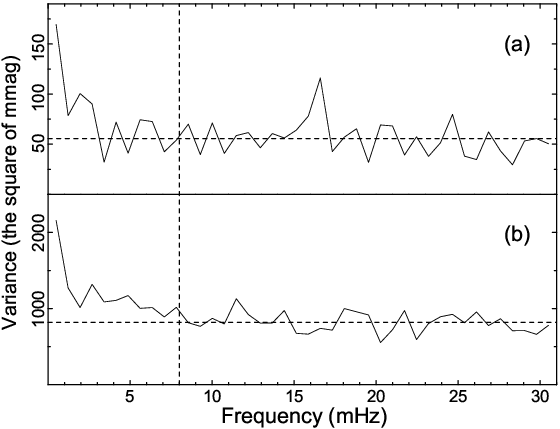}
\caption{Averaged power spectra of short segments of the light curves obtained during the bright (a) and faint (b) states of J2321. The vertical dashed line indicates the approximate limit of 8~mHz for the red noise seen in frame (b). The dashed lines show the white noise levels, which were calculated by averaging the power spectra over a frequency range from 8 to 31~mHz for each of the two cases.}
\label{fig7}
\end{figure}

However, except for the lowest frequencies, the power spectra of the longest light curve segments did not show an increase in power at low frequencies, which would be expected if red noise were present due to the flickering of the star. To enhance the detection of this red noise, I extracted short segments from the light curves of J2321 outside of eclipses. These segments consisted of 256 points and lasted approximately 1.1~hr. In total, I analysed 4 segments when J2321 was in the bright state and 12 segments when it was in the faint state. To eliminate low-frequency variations that were not due to flickering, I used a linear fit to remove any changes on the hourly time scale. Figure~\ref{fig7} illustrates the averaged power spectra of these segments in the two brightness states of J2321. To improve the visibility of the red noise, random fluctuations in these power spectra were smoothed by averaging every three frequency components. As seen, the red noise is present at frequencies up to 3~mHz in the bright state of J2321 (Figure~\ref{fig7}a) and 8~mHz in its faint state (Figure~\ref{fig7}b).

To analyse the flickering in the light curves of J2321, I used Parseval's theorem. This allowed me to relate the noise in the time domain to the noise in the frequency domain, as outlined in \cite{kozhevnikoviz00}. The total variance in a light curve can be thought of as the sum of the variances due to white and red noise: $\sigma^2_{\rm total} = \sigma^2_{\rm white} + \sigma^2_{\rm red}$. Using the Fourier power spectrum for each light curve segment, I calculated the variances caused by total and white noise. The variance caused by red noise was determined as the difference between these two variances. The average total noise, $\sigma_{\rm total}$, was 87$\pm$8 and 335$\pm$19~mmag in the bright and faint states of J2321, respectively. These values agreed with those obtained from light curves directly. The average white noise, $\sigma_{\rm white}$, was 83$\pm$8 and 316$\pm$19~mmag in the bright and faint states of J2321, respectively. These values were close to the photon noises calculated from photon counts. The average red noise, $\sigma_{\rm red}$, was $25\pm2$ and $105\pm10$~mmag in the bright and faint states of J2321, respectively. These average red noises are statistically significant, as they are 10--12 times larger than their standard deviations. Despite the high level of photon noise in the faint state, the red noise can still be easily detected using multiple short segments of light curves.

Although atmospheric interference can contribute to red noise, the atmospheric red noise in our three-channel photometer is relatively low, with a maximum value of 2.7~mmag. In addition, the frequencies of the red noise caused by the atmosphere does not exceed 1.7~mHz \citep[see table 1 and fig. 2 in][]{kozhevnikoviz00}. Furthermore, the red noise caused by the atmosphere should not depend on the brightness of a star. Therefore, I conclude that the red noise observed in the power spectra of J2321 (Figure~\ref{fig7}) is most likely caused by the flickering of the star and not by atmospheric effects. The variance in the light curves of J2321 attributed to red noise, $\sigma^2_{red}$, represents the total flickering power spread across all flickering frequencies. Thus, Fourier analysis of the data revealed that, based on the measured $\sigma_{red}$ values, the flickering power increased by a factor of 18 as the J2321 became fainter.

\section{Discussion}\label{sec4}


I conducted photometric observations of the poorly studied cataclysmic variable J2321 and discovered eclipses. The long-term light curve of J2321 from the ZTF survey showed about a dozen points that were 1.5~mag fainter than other points, indicating eclipses, but nobody had previously tried to determine the period for these points or prove the existence of eclipses in J2321. Using my own photometric data, I was able to identify the eclipses for the first time in J2321.

I observed a total of nine eclipses. By analysing the light curves containing these eclipses, I accurately measured the orbital period of J2321 to be 0.212\,371\,533$\pm$0.000\,000\,069 days. This high precision was made possible thanks to the extensive coverage of my observations, which spanned twelve months.
 
The average depth of eclipses observed in J2321 is 1.36$\pm$0.10~mag, and the profiles of the eclipses follow a Gaussian curve. Based on the short orbital period and the observed eclipse profiles characteristic of CVs, I conclude that J2321 is a genuine CV. In addition, my photometric measurements revealed flickering in J2321, which provides further support for this conclusion.

The high precision of the orbital period allowed me to create a reliable eclipse ephemeris with a long-term validity of approximately four centuries. This ephemeris can be used for future studies of changes in the orbital period, which may be caused by solar-like activity cycles of the red dwarf \citep[e.g.,][]{rubenstein91, applegate92} or the presence of a massive planet orbiting the centre of gravity at J2321 \citep[e.g.,][]{beuermann11, bruch14}. As an eclipsing binary system, J2321 is well suited for determining the masses of its stellar components, as eclipses provide information on the inclination of the orbit. The orbital phases derived from Ephemeris~\ref{ephemeris1} can be used to facilitate further measurements of radial velocity needed to determine masses \citep{hellier01}.



\begin{figure}[t]
\centering
\includegraphics[width=84mm]{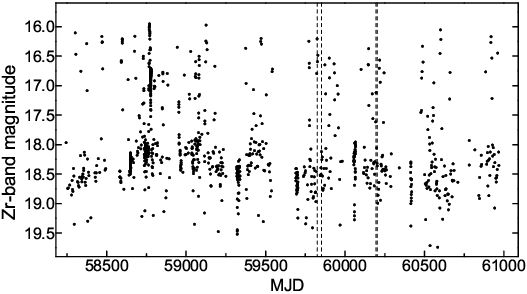}
\caption{Long-term light curve of J2321, obtained during the ZTF survey. The dashed lines represent the times of my observations}
\label{fig8}
\end{figure}

During my observations, the brightness of J2321 outside of eclipses varied between 15.61$\pm$0.02 and 18.04$\pm$0.05~mag, which is lower than typical dwarf nova outburst amplitudes. Moreover, the eclipse profile in J2321 does not change from a V-shape to a U-shape as the brightness decreased (see Figure~\ref{fig6}). This is in contrast to the behavior of dwarf novae during their outburst cycles (see fig.~3 in \citealt{baptista00}). However, the analysis of the data showed that the flickering power significantly increased in the faint state of J2321, similar to what is observed in dwarf novae after outbursts when they become the faintest \citep{bruch21, bruch22}.

As mentioned, the ZTF survey has long-term light curves of J2321 that can help us to understand its nature. The longest $Zr$-band light curve, shown in Figure 8, covers a period of 7.5 years and displays 28 brightness dips reaching magnitudes between 19.2 and 19.8. These dips are identified as eclipses based on Ephemeris~\ref{ephemeris1}. In addition, this light curve shows many points that are brighter by about 2~mag compared to the normal brightness of 18--18.5~ mag, indicating outbursts.

Considering all the light curves from the ZTF survey, I identified 44 individual outbursts consisting of two or more points. The average outburst amplitudes in the $Zr$ and $Zg$ bands are 1.93$\pm$0.07 and 2.39$\pm$0.07~mag, respectively. The average time interval between these outbursts is 37$\pm2$ days.  Based on the Kukarkin-Parenago relationship for dwarf novae, we would expect the amplitude of outbursts at these time intervals to be between 2.9 and 4.7~mag in the $V$ band \citep[see eq. 3.1 in][]{warner95}. However, the observed amplitudes of the outbursts are significantly lower than this prediction. On the other hand, they are higher than the amplitudes of stunted outbursts in nova-like variables, which typically do not exceed 1~mag in amplitude \citep{honeycutt01}. 

To determine whether J2321 is a dwarf nova or a nova-like variable, I used the long-term light curve from the $Gaia$ mission (\url{https://vizier.cds.unistra.fr/viz-bin/VizieR-3?-source=I/355/epphot}) to calculate the time-averaged absolute magnitude of its disc. This was made possible by measuring the parallax and interstellar extinction of J2321. Both of these were provided by the $Gaia$ mission (\url{http://vizier.nao.ac.jp/viz-bin/VizieR-3?-source=I/355/gaiadr3}). In addition, the detected eclipses allowed me to estimate the orbital inclination and adjust the disc brightness to the standard inclination.

The long-term light curve of J2321 from the $Gaia$ mission consists of 56 data points, each with both $G$-band magnitude and $BP-RP$ colour index. It covers a period of 2.7 years. The average $G$-band magnitude was found to be 18.23$\pm$0.11~mag, and the average $BP-RP$ colour index was 1.12$\pm$0.17~mag. Due to smeared eclipses during data collection in the $Gaia$ mission, eclipses cannot be directly removed from the light curve. However, by removing the eclipses from the folded light curve (which is shown in Figure 5), it was determined that the eclipse reduced the average magnitude by only 0.08~mag, resulting in an average $G$-band magnitude of 18.15$\pm$0.11 for J2321 after removal.


Using the parallax and interstellar extinction values from the Gaia DR3 catalogue, Plx = 0.818$\pm$0.118~mas and AG = 0.68$\pm$0.02~mag, the average absolute $G$-band magnitude for J2321 is calculated to be 7.03$\pm$0.34~mag. To convert the $G$-band magnitude to the $V$ band, I used 155 stars from the Hipparcos catalogue (\url{https://vizier.cds.unistra.fr/viz-bin/VizieR-3?-source=I/239/hip_main}) and the photometry obtained by the Gaia mission. I derived a relationship between these two bands: $V-G = 0.015 + 0.052(BP-RP) + 0.1463(BP-RP)^2 + 0.00773(BP-RP)^3$, with a typical scatter of $\pm$0.05~mag. Using this relationship, the average absolute $V$-band magnitude of J2321 was calculated to be 7.30$\pm$0.35~mag. Using eq. 2.102 and fig.~2.46 of \citet{warner95}, I calculated the $M_{\rm V}$ of the red dwarf to be 8.85$\pm$0.50~mag. After removing the light from the red dwarf, the average absolute $V$-band magnitude of the disc is found to be 7.60$\pm$0.35~mag.

The apparent absolute magnitude of the disc must be corrected for the inclination of the orbit. By utilizing the ratio of the masses of the red and white dwarfs and the duration of the eclipse of the white dwarf, it is possible to estimate the inclination of the orbit, as illustrated in fig.~2 in \citet{horne85}. Even if the exact duration of the eclipse of the white dwarf is unknown, the eclipse width at half-depth can still provide a reasonable estimate (\citealt{warner95}, Chapter 2.6.2). The maximum mass ratio for stable mass transfer is 2/3 \citep{warner95}. To determine the minimum mass ratio, I assume that the mass of the white dwarf is 1.44 solar masses (Chandrasekhar limit) and use the orbital period of J2321 and eq. 2.100 from \citet{warner95} to calculate the mass of the red dwarf, yielding a value of $0.50\pm0.02 M_{\rm Sun}$ and a minimum mass ratio of 0.35. The average eclipse width at half-depth is 0.0512(15) phases (Figure~\ref{fig6}c). The range of possible mass ratios and the measured eclipse width at half-depth correspond to a range of potential orbital inclinations between $72.8^\circ$ and $76.3^\circ$. This follows from fig.~2 in \citet{horne85}. Based on eq. 2.63 in \citet{warner95}, this range of inclinations corresponds to a range of disc corrections between 0.92 and 1.23~mag. After applying these corrections, the average absolute $V$-band magnitude of the disc adjusted to the standard inclination varies between 6.37$\pm$0.35 to 6.68$\pm$0.35~mag.

In fig.~3.9 of \citet{warner95}, the solid curve represents the time-averaged absolute $V$-band magnitude of discs corresponding to the critical mass transfer rate. This rate is below which discs are expected to become thermally unstable, leading to dwarf nova outbursts. The curve intersects with a vertical line corresponding to the orbital period of J2321 at 5.4~mag. The range of the possible average absolute $V$-band magnitudes for the disc in J2321 is entirely below this curve. The upper limit is 1.0~mag below this curve and the lower limit is 1.3~mag below this curve. Such deviations from this curve are typical of dwarf novae, as seen in fig.~3.9 of \citet{warner95}. Although J2321 experiences insufficiently large outbursts, which are not consistent with the Kukarkin-Parenago relationship, its low average disc brightness suggests that it should still be classified as a dwarf nova. This conclusion is supported by the increasing flickering power during the low-brightness state of J2321.

\section{Conclusion}\label{sec5}

I conducted photometric observations of the poorly studied cataclysmic variable J2321 and, for the first time, discovered eclipses in this star. The short orbital period, the shape of the eclipses, and the detected flickering all confirm that J2321 is a genuine cataclysmic variable. After careful analysis of my photometric data collected over eight nights spanning twelve months, I obtained the following results:

\begin{enumerate}

\item 
Due to the long time span of observations and the large number of recorded eclipses, I was able to accurately determine the orbital period of the system: $P_{orb} = 0.212\,371\,533\pm0.000\,000\,069$~days. 

\item  The high precision of the orbital period allowed me to create a reliable eclipse ephemeris with a long-term validity of approximately four centuries. This ephemeris can be used for future studies of changes in the orbital period and to facilitate measurements of radial velocity, which are necessary to determine the masses of stellar components.

\item During my observations, the brightness of J2321 outside of eclipses varied between 15.61$\pm$0.02 and 18.04$\pm$0.05~mag, which is lower than typical dwarf nova outburst amplitudes.

\item The eclipses had smooth Gaussian shapes that were the same in both the bright and faint states of J2321. The depths of the eclipses increased from 1.18$\pm$0.02 to 1.51$\pm$0.05~mag as J2321 became fainter. The average depth of the eclipses was 1.36$\pm$0.03~mag. In contrast, the width of the eclipses at half-depth did not change significantly between the two brightness states of the star. On average, it was 0.0512(15) phases wide.

\item I detected the flickering in J2321. The Fourier analysis of the noise in the out-of-eclipse light curves revealed that the flickering power increased by a factor of 18 as J2321 became fainter.

\item Examining the light curves from the ZTF survey, I found that the amplitudes of the outbursts in J2321 were not sufficiently large and did not follow the Kukarkin-Parenago relationship for dwarf novae. However, the increase in flickering power during the faint state of J2321 detected by me was typical of dwarf nova behaviour.

\item Employing the average eclipse width at half-depth, I calculated the potential range of orbital inclinations to be between $72.8^\circ$--$76.3^\circ$. The range of absolute $V$-band magnitudes of the disc in J2321 was found to be between 6.37$\pm$0.35 to 6.68$\pm$0.35~mag. For a CV with an orbital period of 5.1~hr, the average absolute $V$-band magnitude at which the disc becomes thermally unstable is 5.4~mag. Since the average disc brightness in J2321 is fainter by 1.0--1.3~mag than the critical brightness, this CV should be classified as a dwarf nova, despite the fact that it exhibits insufficiently large outburst amplitudes.


\end{enumerate}

\section*{Acknowledgments}

This work was carried out with the financial support of the Ministry of Science and Higher Education of the Russian Federation (theme No FEUZ-2026-0012). The author is grateful to the ZTF team for making their data available to the public. This work has made use of NASA's Astrophysics Data System Bibliographic Services, the VizeR catalogue access tool \citep{ochsenbein00} and the Aladin sky atlas developed at CDS, Strasbourg observatory, France \citep{bonnarel00, boch14}. This work has made use of data from the European Space Agency (ESA) mission $Gaia$ (\url{https://www.cosmos.esa.int/gaia}), processed by the $Gaia$ Data Processing and Analysis Consortium (DPAC, \url{https://www.cosmos.esa.int/web/gaia/dpac/consortium}). Funding for the DPAC has been provided by national institutions, in particular the institutions participating in the $Gaia$ Multilateral Agreement.

\subsection*{Author contributions}

I, Valerij P. Kozhevnikov, am the only author of the manuscript. All content of the manuscript was prepared by me.

\subsection*{Funding}

The author has nothing to report.

\subsection*{Conflict of interest}

The author declares no conflicts of interest.

\subsection*{Data Availability Statement}

The data that support the findings of this study are openly available in
publication at \url{https://www.researchgate.net/publication/415561228}, reference number 415561228.

\bibliography{kozhevnikov_ms}%



\end{document}